\documentclass[a4paper]{article}

\usepackage{authblk}
\usepackage[utf8]{inputenc}
\usepackage[T1]{fontenc}
\usepackage{pslatex}
\usepackage[english]{babel}
\usepackage{fancyhdr}
\usepackage{hyperref}
\usepackage{multirow}
\usepackage{amsmath,amssymb,amsfonts}
\usepackage{graphicx}
\usepackage{textcomp}
\usepackage{xcolor}
\usepackage{url}
\usepackage{hyperref}
\usepackage{microtype}
\usepackage{subcaption}
\usepackage{tikz}
\usetikzlibrary{positioning}
\usetikzlibrary{decorations.pathreplacing}
\usepackage{cite}
\usepackage{wasysym}
\usepackage{booktabs}
\usepackage{paralist}
\usepackage{hyperref}
\usepackage{nomencl}
\usepackage{listings}
\usepackage{algpseudocode}
\usepackage{algorithmicx}

\algblockdefx[Foreach]{Foreach}{EndForeach}[1]{\textbf{for each} #1 \textbf{do}}{\textbf{end for}}

\newcommand{\N}{\mathbb{N}}
\newcommand{\F}{\mathbb{F}}

\providecommand{\keywords}[1]{\textbf{\textit{Keywords }} #1}

\begin{document}

\title{A Survey of Metaheuristic Algorithms for the Design of Cryptographic Boolean Functions}

\author[1]{Marko Djurasevic}
\author[1]{Domagoj Jakobovic}
\author[2]{Luca Mariot}
\author[3]{Stjepan Picek}

\affil[1]{{\normalsize University of Zagreb, 10000 Zagreb, Croatia} \\

    {\small \texttt{\{marko.durasevic, domagoj.jakobovic\}@fer.hr}}}

\affil[2]{{\normalsize Semantics, Cybersecurity \& Services (SCS), University of Twente, Drienerlolaan 5, 7522NB Enschede, The Netherlands} \\

{\small \texttt{l.mariot@utwente.nl}}}

\affil[3]{{\normalsize Digital Security Group, Radboud University, Postbus 9010, 6500 GL Nijmegen, The Netherlands} \\
	
	{\small \texttt{stjepan.picek@ru.nl}}}
	
\maketitle

\begin{abstract}
Boolean functions are mathematical objects used in diverse domains and have been actively researched for several decades already. One domain where Boolean functions play an important role is cryptography. There, the plethora of settings one should consider and cryptographic properties that need to be fulfilled makes the search for new Boolean functions still a very active domain. There are several options to construct appropriate Boolean functions: algebraic constructions, random search, and metaheuristics. In this work, we concentrate on metaheuristic approaches and examine the related works appearing in the last 25 years. To the best of our knowledge, this is the first survey work on this topic. Additionally, we provide a new taxonomy of related works and discuss the results obtained. Finally, we finish this survey with potential future research directions.
\end{abstract}

\keywords{Boolean functions, Cryptography, Metaheuristics, Representations, Truth Table, Walsh-Hadamard Transform}

\section{Introduction}
\label{sec:intro}

Boolean functions represent mathematical objects with diverse applications.
For instance, in combinatorial designs, Boolean functions are used to construct Hadamard matrices~\cite{Rothaus} and strongly regular graphs~\cite{954512}.
In coding theory, every binary unrestricted code of length $2^n$ can be interpreted as a set of Boolean functions. In sequences, bent sequences constructed using bent Boolean functions have the lowest value of mutual correlations and autocorrelations, and they are used in communication systems with multiple access~\cite{1056589}. In telecommunications, bent Boolean functions are used in CDMA networks~\cite{Paterson}. In cryptography, Boolean functions are used in stream and block ciphers as the source of nonlinearity~\cite{4036240}. Furthermore, they are used in fully homomorphic encryption~\cite{Meaux2016} and in the design of hash functions~\cite{Zheng1993}.

Due to their widespread use, the body of works considering their various aspects is rich and spans several decades. Many of those works consider how to construct Boolean functions with specific properties. Considering the methodology, such works can be classified into algebraic constructions, random search, and metaheuristics. 
While this division makes sense from an intuitive point of view, different characteristics of those approaches make any comparison difficult for the following reasons:
\begin{compactenum}
\item Algebraic constructions work for many different Boolean function sizes but always produce the same Boolean functions when adopting the same starting conditions. Finding new constructions can be far from trivial.
\item Random search is easy to run and will, in principle, output many different Boolean functions. Still, as the required size of the Boolean function increases, the resulting properties become suboptimal.
\item Metaheuristics can produce many different Boolean functions that commonly have better properties than those obtained with random search (rivaling those obtained with algebraic constructions). Still, such techniques will struggle to produce high-quality solutions for larger Boolean function sizes. Next, one needs to devise an appropriate objective function to guide the metaheuristic search, which is more difficult than simply running a random search but significantly simpler than devising a new algebraic construction. Finally, it is necessary to run a new search for each considered Boolean function size.
\item Potentially, one could also devise a construction technique than combines different approaches. For instance, using algebraic constructions to find Boolean functions with good properties and then using those functions to optimize upon with metaheuristics. Additionally, one commonly uses random search to construct the initial solutions for metaheuristics. 
\end{compactenum} 

Clearly, metaheuristics have some obvious advantages over other construction techniques. As such, various metaheuristic techniques have been tested over the years concerning the design of Boolean functions with specific properties and dimensions. 
While most of those works report favorable results, there are still many challenges. One of the big challenges is the lack of a comprehensive overview of the contributions that have been achieved.
In this work, we systematically survey the work done and the open challenges. Moreover, we suggest a new division of metaheuristic approaches into 1) direct design and 2) metaheuristic-assisted design of Boolean functions. With this division, we argue to capture the most important differences in the applied metaheuristic techniques and allow a better comparison of the obtained solutions. Next, we propose to further divide works in the direct design of Boolean functions based on the representation of Boolean functions into truth table-based approaches and Walsh-Hadamard-based approaches. Concerning the part on metaheuristic-assisted design, we classify relevant research into the evolution of algebraic constructions and the optimization of combinatorial objects related to Boolean functions, such as orthogonal arrays.

The rest of this paper is organized as follows. Section~\ref{sec:background} recalls basic background notions related to Boolean functions and gives an overview of the most common metaheuristic algorithms used in the literature to optimize their cryptographic properties. Section~\ref{sec:direct_evolution} surveys the works belonging to the first main category of our proposed taxonomy, namely the direct design of Boolean functions via metaheuristics. Section~\ref{sec:assisted_construction} gives an overview of the field of metaheuristic-assisted design of Boolean functions. Finally, Section~\ref{sec:directions} points out several directions for future research on the subject, and Section~\ref{sec:conclusions} concludes the paper.

\section{Background}
\label{sec:background}

Let $\F_2 = \{0,1\}$ be the finite field with two elements, with sum and multiplication of $a,b \in \F_2$ respectively corresponding to the XOR (denoted by $\oplus$) and logical AND (denoted by concatenation) of $a$ and $b$. Given any positive integer $n \in \N$, we denote by $\F_2^n$ the set of all $n$-tuples of elements in $\F_2$, which we endow with a vector space structure. In particular, the sum of two vectors $x, y \in \F_2^n$ amounts to their bitwise XOR, while the multiplication of $x \in \F_2^n$ with a scalar $a \in \F_2$ is the logical AND of $a$ with each coordinate of $x$. Slightly abusing notation, we still denote vector sum and multiplication by a scalar, respectively by $\oplus$ and concatenation. We also consider $\F_2^n$ as a metric space by defining the Hamming distance $d_H(x,y)$ of two vectors $x, y \in \F_2^n$ as the number of coordinates where $x$ and $y$ differ. The support of a vector $x \in \F_2^n$ is the subset of positions where $x$ is nonzero, i.e., $supp(x) = \{i: x_i \neq 0 \}$, with the Hamming weight $w_H(x)$ of $x$ being the cardinality of its support. Finally, we endow the vector space $\F_2^n$ with the inner product $\cdot : \F_2^n \times \F_2^n \to \F_2$, defined for all $x, y \in \F_2^n$ as $x \cdot y = \bigoplus_{i=1}^{n} x_{i}y_{i}$. Thus, the inner product between $x$ and $y$ is the sum (over $\F_2$) of the pointwise multiplications of $x_i$ and $y_i$.

\subsection{Boolean Functions and their Representations}
\label{subsec:bf-rep}

An $n$-variable Boolean function is any mapping of the form $f: \F_2^n \to \F_2$, and it can be uniquely represented by the truth table (lookup table), which is the list of pairs of function inputs $x \in \mathbb F_2^n$ and corresponding function values $f(x)$. The value vector is the binary vector $\Omega_f$ composed of all $f(x)$, with $ x \in \mathbb{F}_2^n$, where some total order has been fixed on $\mathbb{F}_2^n$ (most commonly, the lexicographic order). The size of the value vector equals $2^n$, thus we have that $\Omega_f \in \F_2^{2^n}$. If we denote the set of all $n$-variables Boolean functions by $\mathcal{B}_n = \{f: \F_2^n \to \F_2\}$, it follows that the size of $\mathcal{B}_n$ is $\#\mathcal{B}_n = 2^{2^n}$, i.e., it grows super-exponentially in $n$ (see Table~\ref{tab:nr_boolean} for exact and approximate values of $\#\mathcal{B}_n$). 
\begin{table}
\small
  \centering
  \caption{Exact ($\#\mathcal{B}_n$) and approximate ($\approx\mathcal{B}_n$) numbers of $n$-variable Boolean functions.}
  \label{tab:nr_boolean}
  \begin{tabular}{cccccc}
    \toprule
    $n$ & $3$ & $4$ & $5$ & $6$ & $7$ \\ \midrule
    $\#\mathcal{B}_n$ & $256$ & $65536$  & $2^{32}$ & $2^{64}$ & $2^{128}$\\
    $\approx\mathcal{B}_n$ & 256 & 65536 & $4.29 \cdot 10^9$ & $1.84 \cdot 10^{19}$ & $3.40 \cdot 10^{38}$\\ 
    \midrule
    $n$ & $8$  & $9$ & $10$ & $11$ & $12$\\ \midrule
    $\#\mathcal{B}_n$ & $2^{256}$  & $2^{512}$ & $2^{1024}$ & $2^{2048}$ & $2^{4096}$\\
    $\approx\mathcal{B}_n$ & $1.16 \cdot 10^{77}$ & $1.34 \cdot 10^{154}$ & $1.80 \cdot 10^{308}$ & $3.23 \cdot 10^{617}$ & $1.04 \cdot 10^{1234}$ \\ 
    \bottomrule
  \end{tabular}
\end{table}
The support and the Hamming weight of a Boolean function $f: \F_2^n \to \F_2$ correspond respectively to the support and Hamming weight of its value vector $\Omega_f$. While the truth table representation is ``human-friendly'', not much can be directly deduced from it, except for the Hamming weight.

A second unique representation of a Boolean function $f$ on $\F_{2}^{n}$ is the Algebraic Normal Form (ANF). Remarking that $x^2 = x$ for all $x \in \F_2$, the ANF of $f: \F_2^n \to \F_2$ is the multivariate polynomial in the quotient ring $\F_{2}\left[x_{0},..., x_{n-1}\right]/(x_{0}^{2} \oplus x_{0},..., x_{n - 1}^{2} \oplus x_{n - 1})$, defined as:
\begin{equation}
\label{eq:anf}
f(x) = \bigoplus_{\substack{a \in \F_{2}^{n}}} h(a)\cdot x^{a},
\end{equation}  
\noindent
where $x^a = (x_0^{a_0},\cdots, x_{n-1}^{a_{n-1}})$, and $h(a) \in \F_2$ is given by the M\"{o}bius transform:
\begin{equation}
\label{eq:moebius}
h(a)= \bigoplus_{\substack{x \preceq a}}  f(x), \text{ for any } a \in \F_2^n.
\end{equation}
Here, $x \preceq a$ means that $a$ covers $x$ (alternatively, $x$ precedes $a$), which means that $x_i \leq a_i$, for all $i \in \left\lbrace 0, \ldots, n-1 \right\rbrace$. The M\"{obius} transform is an involution: one can retrieve the value of the truth table of $f$ by swapping $h(a)$ with $f(x)$ and $a$ with $x$ in Eq.~\eqref{eq:anf}. The algebraic degree of $f$ is the size of the largest nonzero monomial in the ANF of $f$, formally defined as:
\begin{equation}
\label{eq:deg}
deg(f) = \max_{a: h(a) \neq 0} \left\{ w_H(a) \right\}.
\end{equation}
Functions of degree at most 1 are also called affine. An affine function $f$ is called linear if $h(\underbar{0}) = 0$, i.e., if its ANF does not have a constant term. It is easy to see that the ANF of a linear function corresponds to the inner product $a \cdot x$ between a vector $a \in \F_2^n$ and the input vector $x \in \F_2^n$.

Another way to uniquely represent a Boolean function $f: \F_2^n \to \F_2$ is the Walsh-Hadamard Transform (WHT) $W_f: \F_2^n \to \mathbb{Z}$, defined for all $a \in \F_2^n$ as:
\begin{equation}
\label{eq:wht}
W_{f} (a) = \sum\limits_{x \in \F_{2}^{n}} (-1)^{f(x) \oplus a\cdot x}.
\end{equation}
In particular, the coefficient $W_f(a)$ measures the correlation between $f$ and the linear function $a \cdot x$. The multiset of all coefficients $W_f(a)$ for $a \in \F_2^n$ is also called the Walsh-Hadamard Spectrum (WHS) of $f$. The WHT is very useful in cryptography as many cryptographic properties of Boolean functions can be characterized through it. Contrarily to the ANF, the WHT is not an involution since it maps the set $\F_2^{2^n}$ to $\mathbb{Z}^{2^n}$. The mapping $W_f$ is however injective, from which it follows that the spectrum of a Boolean function $f$ uniquely identifies $f$. In particular, one can retrieve the truth table of $f$ from its WHS by using the Inverse Walsh-Hadamard Transform, which has the same structure of Eq.~\eqref{eq:wht}, except that $x$ and $(-1)^{f(x)}$ are replaced respectively by $a$, and $W_f(a)$, and the sum is normalized by a $2^{-n}$ factor.

From a computational complexity point of view, a naive algorithm to compute the ANF or the WHT of a $n$-variable Boolean function requires $\mathcal{O}(2^{2n})$ steps since one needs to loop over all possible values $a \in \F_2^n$, and for each of them, the sum ranges again in $\F_2^n$. However, there exist more efficient divide-and-conquer butterfly algorithms, namely the Fast M\"{obius} Transform and the Fast Walsh-Hadamard Transform that requires only $\mathcal{O}(n2^n)$ steps. Details of such algorithms can be found in~\cite{carlet_2021}.

\subsection{Cryptographic Properties and Bounds}
\label{subsec:prop}

Boolean functions used in the stream and block ciphers model must fulfill several cryptographic criteria. Each criterion is geared towards a particular cryptanalytic attack: the rationale is that if a Boolean function satisfies it, then mounting the corresponding attack becomes computationally unfeasible for the attacker. 

Here, we only discuss the most often considered properties in the works addressing the construction of Boolean functions with metaheuristics. For more information on the attacks they protect from and other cryptographic properties not covered here that have been less frequently considered with metaheuristics (such as propagation criteria and algebraic immunity), we refer the reader to~\cite{carlet_2021}.

\paragraph{Balancedness} The Hamming weight of a Boolean function $f:\F_2^n \to \F_2$ is a basic cryptographic property, which indicates the output bias of the function. In particular, the Boolean function should be balanced, namely $w_H(f) = 2^{n-1}$. This means that the truth table of $f$ is composed of an equal number of zeros and ones. This property can also be expressed in terms of the WHT as follows: $f$ is balanced if and only if $W_f(\underbar{0})=0$.

\paragraph{Algebraic Degree} We defined the algebraic degree of $f$ in Section~\ref{subsec:bf-rep} as the size of the largest occurring monomial in the ANF of $f$. As a cryptographic criterion, the algebraic degree of $f$ should be as high as possible.

\paragraph{Nonlinearity} The minimum Hamming distance between a Boolean function $f$ and all affine functions is called the nonlinearity of $f$, denoted by $nl_{f}$. This property can be characterized in terms of the Walsh-Hadamard coefficients as follows:
\begin{equation}
\label{eq:nonlinearity}
nl_{f} = 2^{n - 1} - \frac{1}{2}\max_{a \in \F_{2}^{n}} |W_{f}(a)|.
\end{equation}
As a cryptographic property, the nonlinearity of $f$ should be as high as possible. According to Eq.~\eqref{eq:nonlinearity}, this happens if the largest Walsh-Hadamard coefficient in absolute value is as low as possible. Parseval's identity states that the sum of the squared Walsh-Hadamard coefficients is constant for all $n$-variable Boolean functions, and it equals $2^{2n}$. This result allows for deriving the following inequality, known as the covering radius bound:
\begin{equation}
\label{eq_boolean_covering}
    nl_{f} \leq 2^{n-1}-2^{n/2-1}.
\end{equation}
A Boolean function can be considered highly nonlinear if its nonlinearity is close to the covering radius bound in its class. In particular, the functions whose nonlinearity equals the maximal value $2^{n-1}-2^{n/2-1}$ are called bent. Bent functions exist only for even values of $n$, as each WHT coefficient must be equal to $\pm 2^{\frac{n}{2}}$. Consequently, bent functions are not balanced since $W_f(\underbar{0}) \neq 0$. When $n$ is odd, the bound in Eq.~\eqref{eq_boolean_covering} cannot be tight. In this case the maximum nonlinearity is between $2^{n-1} - 2^{\frac{n-1}{2}}$ and $2^{n-2} - 2^{n/2-2}$~\cite{carlet_2021}. The former value is also called the quadratic bound since it is the maximum nonlinearity reachable by Boolean functions with algebraic degree $2$.

\paragraph{Correlation Immunity and Resiliency}
A Boolean function $f: \F_2^n \to \F_2$ is $t$-th order correlation immune, with $1 \le t \le n$, if its output distribution does not change by fixing at most $t$ input variables. This property is characterized using the Walsh-Hadamard Transform as follows: $f$ is $t$-th order correlation immune if and only if $W_f(a) = 0$ for all $a \in \F_2^n$ such that $1 \le w_H(a) \le t$. As a cryptographic criterion, a Boolean function should be correlation immune of a high order. Further, a function is $t$-resilient if it is balanced and $t$-th order correlation immune. This means that any restriction of the function obtained by fixing at most $t$ coordinates is a balanced function. The order $t$ of correlation immunity (respectively, resiliency) induces a trade-off with the algebraic degree and nonlinearity of a function. More precisely, Siegenthaler's bound states that $d \le n - t$ (respectively, $d \le n-t-1$), where $d$ is the algebraic degree of the function; moreover, a consequence of Sarkar-Maitra's divisibility bound is that $nl_f \le 2^{n-1} - 2^{t}$ (respectively, $nl_f \le 2^{n-1}-2^{t+1}$).

\subsection{Metaheuristics}

Heuristics are algorithms that find good solutions to large-size problem instances. 
In general, they do not have an approximation guarantee on the obtained solutions~\cite{Talbi}. 
Alternatively, heuristics can be defined as parts of an optimization algorithm. 
In that role, heuristics use the information currently gathered by the algorithm to help decide which solution candidate should be tested next or how the next solution can be produced.

Heuristic algorithms can be further divided into problem-specific heuristics and metaheuristics.
Problem-specific heuristics are methods that are tailor-made to solve a specific problem.
Metaheuristics, in the original definition, represent solution methods that orchestrate an interaction between local improvement procedures and higher-level strategies to create a process capable of escaping from local optima and performing a robust search of a solution space~\cite{meta}. 
Alternatively, metaheuristics can be defined as general-purpose algorithms that can be applied to solve almost any optimization problem~\cite{Talbi}.
One can follow many criteria to classify metaheuristics, but we divide them into single-solution-based metaheuristics and population-based heuristics~\cite{Talbi}.
Single-solution-based metaheuristics manipulate and transform a single solution during the search, as in the case of algorithms like Local Search (LS) or Simulated Annealing (SA).
Population-based metaheuristics work on a population of solutions, e.g., Evolutionary Algorithms (EAs), Particle Swarm Optimization (PSO), and Artificial Immune Systems (AIS).

\subsubsection{Local Search}

Local Search (LS) is possibly the simplest metaheuristic method that, in each iteration, replaces the current solution with a neighbor that improves the objective function~\cite{Talbi}. 
In each iteration, the algorithm searches a neighborhood $N(t)$ of the current solution and selects a better solution, if one can be found, for the next iteration.
To generate neighboring solutions of the current solution, LS can use various operators, which differ in how they generate the neighborhood. 
Although many strategies can be used to select which neighbor should be selected from $N(t)$ to replace the current solution, such as first improvement, best improvement, or random, all of them lead to the same problem.
LS is inherently a greedy hill-climbing method that will get trapped in the first local optimum it reaches since it only accepts solutions that are better than the current one. 
Therefore, the basic LS method was extended in different ways, which allowed it to escape local optima and achieve better performance on multimodal problems.

\subsubsection{Simulated Annealing}

Simulated Annealing (SA) operates on a single potential solution, which is locally changed in each iteration, and its new fitness value is recorded~\cite{Kirkpatrick1983}. 
The algorithm is inspired by the annealing process of metals, in which a certain material is heated and then gradually cooled to alter its physical properties.
Similar to LS, SA uses neighborhood operators to search for better solutions in the vicinity of the current solution. 
However, in addition to accepting a better neighboring solution, SA can also accept a worse neighbor with a certain probability.
The idea is that the probability of accepting worse solutions gradually decreases, allowing the algorithm to thoroughly explore the search space in earlier iterations and converge to an optimal solution in later iterations.
The acceptance probability of worse solutions is controlled by a parameter denoted as the temperature, with different cooling schedules being used to decrease the temperature during the execution of the algorithm, which also decreases the probability of accepting worse solutions.

\subsubsection{Evolutionary Algorithms}
\label{sec:eas}

Evolutionary algorithms (EAs) are population-based metaheuristic optimization methods that use biology-inspired mechanisms like selection, crossover, and survival of the fittest~\cite{Eiben03}.
The biggest milestone in the development of this area can be traced to the 19th century and Charles Darwin. In 1859, he published a book, ``On the Origin of Species'' where he identified the principles of natural evolution~\cite{Darwin}.
There are many different evolutionary algorithms, but they all share the same general traits.
The usual variation operators are mutation and crossover. Crossover operates over multiple solutions (parents) and combines them into a new solution (child). Mutation works on a single individual and changes parts of it. 
The goal of crossover is to facilitate the exploitation of good properties of existing solutions, whereas mutation increases the diversity and thus stimulates the exploration of new solutions.
The selection operator also represents an important operator used to select solutions that should participate in the crossover or be eliminated from the population. 
Although various selection operators were proposed, their logic is the same in the sense that they foster the choice of better individuals for recombination while giving a larger chance of eliminating worse solutions from the population.

\paragraph{Genetic Algorithm}

Genetic algorithms (GAs) are probabilistic algorithms where search methods model some natural phenomena: genetic inheritance and survival of the fittest.
GAs are a subclass of EAs where the elements of the search space are arrays of elementary types like strings of bits, integers, floating-point values, and permutations~\cite{Eiben03}.

\paragraph{Genetic Programming}

Genetic Programming (GP) belongs to EAs and commonly uses tree data structures that undergo an evolutionary process~\cite{koza}.
Although GP has a history of more than 50 years, its full acceptance is due to the work of Koza at the beginning of the 1990s, when he formalized the idea of employing chromosomes based on tree data structures. 
Since the aim of GP is to generate new programs automatically, each individual in a population represents a computer program. The most common form is a symbolic expression representing a parse tree. A parse tree is an ordered, rooted tree that represents the syntactic structure of a string according to some context-free grammar. 
A tree can represent, e.g., a mathematical expression, a rule set, or a decision tree.
The building elements in a tree-based GP are functions (inner nodes) and terminals (leaves, problem variables). Both functions and terminals are known as primitives. 

\subsubsection{Particle Swarm Optimization}

Particle Swarm Optimization (PSO) is one of the most well-known and successful representatives of swarm intelligence methods~\cite{Talbi}, which model an indirect communication between the individuals in the population as they traverse the search space.
Concretely, PSO is inspired by the social behavior of birds, in which a coordinated behavior can be observed as they migrate in flocks~\cite{Kennedy1995}.
Each solution in PSO, denoted a particle, consists of its current position and velocity, which are updated in each iteration of the algorithm.
The position specifies the current position of the particle in the search space, whereas the velocity represents the direction in which the particle will move in search of the global optimum. 
The velocity is updated as a linear combination of the previous velocity and the relative position of the particle representing the best solution obtained by this particle and the best overall solution obtained by any particle in the swarm.
After updating the velocity, it is used to update the position of each particle by adding it to the current position of the particle.

\subsubsection{Artificial Immune Systems}

Artificial Immune Systems (AIS) is a group of metaheuristics inspired by concepts observed in the immune systems of living beings~\cite{Talbi}.
Although various algorithms inspired by immune systems were proposed, one of the most popular and commonly used algorithms is the Clonal Selection Algorithm (CLONALG)~\cite{Castro2002}. 
In this algorithm, each solution called an antibody, undergoes a cloning process in which a number of clones of each solution are created. 
The number of clones created for each solution is proportional to the affinity (quality) of each antibody, meaning more clones of better solutions will be created. 
After cloning, the hypermutation operator is applied to modify the clones with a certain probability, again based on their affinity.
In this case, the mutation probability is larger for antibodies with a lower affinity, meaning that smaller changes are performed on good solutions.
The new population is created by selecting a certain number of the best-created clones.
However, to foster diversity, a certain number of the worst solutions in the new population are replaced with randomly generated solutions.

\section{Direct Design of Cryptographic Boolean Functions with Metaheuristics}
\label{sec:direct_evolution}

Constructing a Boolean function can be regarded as a \emph{combinatorial optimization problem} where, given the number of variables $n$ and the set of properties to optimize, the goal is to find an $n$-variable Boolean function that satisfies the desired properties.
The first choice one needs to make when optimizing a Boolean function is the representation of candidate solutions.
In most applications, this has also proven to be the most influential factor, while the choice of the actual search algorithm had a smaller impact.
It is important to note that the majority of metaheuristics, both the ones presented in the previous chapter, as well as in general, allow the use of an arbitrary solution representation.
However, a set of appropriate operators required by the selected algorithm must be defined for each encoding.

\subsection{Truth Table-based Representations}
\label{subsec:tt}

In this section, we present the applications that rely on the optimization of the underlying truth table of a Boolean function, as opposed to optimizing the Walsh-Hadamard function representation.
Since most related works consider the truth table representation, we do not describe each paper in detail but concentrate on unifying perspectives.

\subsubsection{Solution Encodings}

\paragraph{Bitstring.}

In the context of Boolean functions, the bitstring representation is the most trivial approach to encoding a solution. 
In this case, the algorithm operates directly on the truth table since a potential solution is encoded as a string of bits with a length of $2^n$ for a Boolean function with $n$ inputs.
This encoding is the most natural for many metaheuristics, such as the genetic algorithm, as the algorithm is based on the genetic code paradigm.
However, the application of this encoding will ultimately depend on the type of the search algorithm.

Single-solution-based search methods need to define at least one neighborhood for the solution encoding, and in this case, there are many possibilities.
The usual neighborhood operators include inverting a single bit, setting or resetting a bit, exchanging two randomly selected bits (swap), inverting a sequence between two-bit positions, mixing the bits between two positions, inserting a bit at a randomly selected position, etc.
These local operators are also commonly used as mutation operators in evolutionary algorithms. 

When only balanced Boolean functions are considered, the set of operators is usually constrained to the ones that preserve that property (such as swap, inversion, or insertion).
Additionally, new operators may be defined for a certain property, such as inverting two random bit positions instead of a single one in the case of balanced functions.

In addition to mutation, evolutionary algorithms also need to define crossover operators, which combine genetic material from (at least) two parent solutions to construct a single-child solution.
Here, the most common choice is the one-point crossover, which randomly selects a breakpoint in the string; the child solution is a concatenation of the bits from the first parent up to the breakpoint and from the second parent onward.
In the case of larger solutions, more breakpoints can be used to promote diversity.
The extreme case is a uniform crossover that randomly selects the source parent after each bit position, which is the most disruptive method. As for mutation, special crossover operators can be considered when the goal is to constrain the search space only to balanced functions. Examples of this approach include uniform crossover augmented with counters that keep track of the multiplicities of 0s and 1s in the child chromosome. When one of the two counters reaches half of the bitstring length, the remaining positions are filled with the complementary value to maintain the balance~\cite{10.1007/BFb0054148,manzoni20}.

Most papers on Boolean functions include the bitstring representation for optimization of Boolean functions, but it is rarely used as a single standalone encoding (see Table~\ref{tab:encodings}).

\paragraph{Integer.}

A more compact way of representing the truth table is by way of grouping substrings of bits into separate integer values.
Indeed, in many implementations, this is the actual way of storing a sequence of bits in memory.
In this case, the truth table is divided into a number of substrings $k$ of the same length $l$, and each substring is represented with an integer value.
For this to be feasible, the truth table size $2^n$ must be divisible by $k$, and the substring length $l$ must be a power of 2; each integer value is then restricted to $[0, 2^l-1]$.

Using a different encoding allows the search algorithm to operate on a different data structure, which is usually (and in the context of evolutionary computation) called the \textit{genotype}.
When a candidate individual is evaluated, its genotype is translated into the \textit{phenotype}, which is the actual representation of the solution of the problem being solved.
In this case, the genotype the algorithm operates on is the sequence of integers, while the phenotype is still in the form of a truth table of the Boolean function.

Several options are available when decoding the genotype to the phenotype: the integer value can be translated to a binary string using natural binary encoding, Gray coding, or some other scheme. 
Further, binary strings representing each integer can be \textit{concatenated} to form a complete truth table. On the other hand, the individual bits can be \textit{distributed} so that bits encoding one integer value are placed $k$ positions apart from each other in the truth table.

Under an integer encoding, the local search (or mutation) and crossover operators will behave differently from the bitstring case, thus forming different neighborhood structures.
The common mutation operators include randomly modifying a single gene (integer value), either in the whole range $[0, 2^l-1]$ or by a smaller value, gene swapping, substring inversion, etc.
Crossover operators are usually based on one-point or multiple-point crossover; however, some operators combine each pair of genes from two parents independently, with the resulting child gene assuming an average of two-parent genes or a random value between the parent genes.

Integer encodings have also been used when restricting the search space to balanced Boolean functions. The main approaches are the map-of-ones and the zero-length encodings~\cite{manzoni20}. In the former, the genotype of a balanced $n$-variable function is defined by a sequence of $2^{n-1}$ integer numbers between $0$ and $2^{n}-1$, which represent the positions of the 1s in the truth table vector of the function. In the latter, the function is instead represented by a sequence of $2^{n-1}+1$ integer numbers. Each number specifies how many consecutive 0s are between two 1s, and balancedness is ensured by requiring that the sum of all numbers in the sequence is $2^{n-1}$. Clearly, in both approaches, special crossover and mutation operators are required to generate offspring chromosomes that preserve the corresponding encodings.

\paragraph{Floating-point.}

The previous encodings are well suited for genotype-agnostic search algorithms, such as GA or local search methods.
Metaheuristics that are defined over the real-valued domain, such as PSO and AIS, can, in principle, be modified to operate on bit strings or integer values as well but are rarely used in optimizing Boolean functions in this form.

To accommodate algorithms operating on the continuous domains, we can take one step further and encode a single integer value as a floating-point (FP) number.
A common practice is to define the genotype as a sequence of floating-point numbers assuming values in $[0, 1]$ or $[-1, 1]$.
Real values are first decoded into corresponding integer values in $[0, 2^l-1]$ and then to the resulting truth table, as in the previous integer encoding.

Since the resolution of the floating point is the highest around 0, a single FP value can represent a relatively large number of truth table bits.
In theory, this could be scaled up to around 48 bits since $2^{-48} \simeq 10^{-16}$, which is the approximate rounding error at magnitude $1$ for 64-bit FP values.
However, usually, a much smaller number of bits has actually been used in practice, with the boundary case being one FP value used to encode just a single bit~\cite{PICEK2017320}.

Using floating-point representation allows the application of many optimization algorithms designed for continuous optimization.
Algorithms of this type may employ highly specific modification operators in the real value domain, resulting in neighborhood structures that may differ substantially from those used in the discrete binary domain.
However, not many papers present examples of this approach, as only one instance of continuous optimization algorithm has been applied to Boolean optimization (Table~\ref{tab:encodings}). As a matter of fact, the only two works we are aware of that use PSO to evolve the cryptographic properties of Boolean functions~\cite{saber06,10.1007/978-3-319-10762-2_80} actually use a \emph{discrete} variant of this metaheuristic, rather than the basic version tailored for continuous search spaces. Furthermore, while the objective (fitness) landscape has been an object of active research in bitstring encoding (e.g., \cite{10.1145/2739480.2754739, JAKOBOVIC2021107327}), we are not aware of such an analysis for the floating-point encoding.

\paragraph{Symbolic.}

Regardless of the application, all optimization methods using direct mapping to truth table-based representation suffer from the same problem: the curse of dimensionality.
Although good results may be obtained for small Boolean function sizes, the efficiency inevitably deteriorates with the increase in the number of variables.
A different approach partially circumvents this problem by relying on a \textit{symbolic} representation of a Boolean function as a genotype.

Evolutionary algorithms that use symbolic solution representation, such as Genetic Programming (GP), are usually applied to this goal. As recalled in Section~\ref{sec:eas}, GP maintains a population of candidate programs for solving a given problem. The programs may take any form, but the most common representation in GP uses a \textit{syntax tree} encoding, where inner tree nodes represent functional elements, and the terminal nodes represent variables or functions without arguments. If the execution of the program does not produce any side effects, then the program is equivalent to a function, and in this case, the GP solves a \textit{symbolic regression} problem.

\begin{figure}
    \centering
    \includegraphics{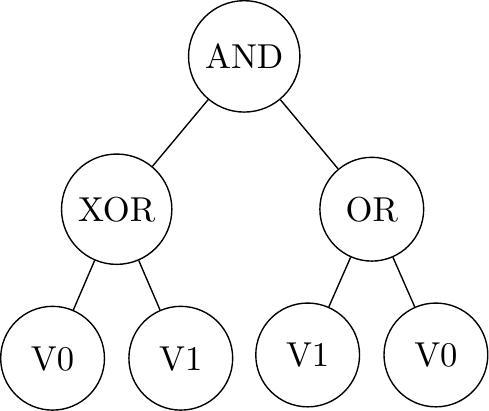}
    \caption{Symbolic representation used by tree-based GP}
    \label{fig:gprepr}
\end{figure}

Symbolic regression may be regarded as equivalent to the problem of evolving (finding) a suitable Boolean function satisfying one or more given properties.
Here, the $n$ Boolean variables present terminal nodes (leaves) that can appear as arguments in a syntax tree.
Commonly used functional elements in this scenario include a number of elementary binary Boolean functions, such as OR, XOR, AND, XNOR, etc.
Apart from these, unary functions such as NOT are also used, as well as non-standard functional elements: AND with one input inverted, IF with three arguments (that evaluates the first argument, returns the second one if the first evaluates to 'true', and the third otherwise), etc.
Figure~\ref{fig:gprepr} shows an example of a symbolic tree used by GP, which represents the Boolean expression (V0 XOR V1) AND (V1 OR V0).

A GP tree thus represents an executable expression that is evaluated as a Boolean function.
Regardless of the tree elements and shape, a candidate tree is evaluated by ``executing" the expression for every possible combination of input Boolean variables and recording its output, thus generating the function truth table as its phenotype.
The conversion process from genotype to phenotype may be time-demanding for a larger number of variables since the same expression needs to be interpreted $2^n$ times.

In GP, various mutation and crossover operators are used; for instance, subtree mutation randomly selects a node within the tree and replaces the selected subtree with a new randomly created one.
Crossover operators may include simple tree crossover, uniform crossover, size fair, one-point, context preserving crossover, etc.~\cite{GPguide}.

\begin{figure}
    \centering
    \includegraphics[width=\textwidth]{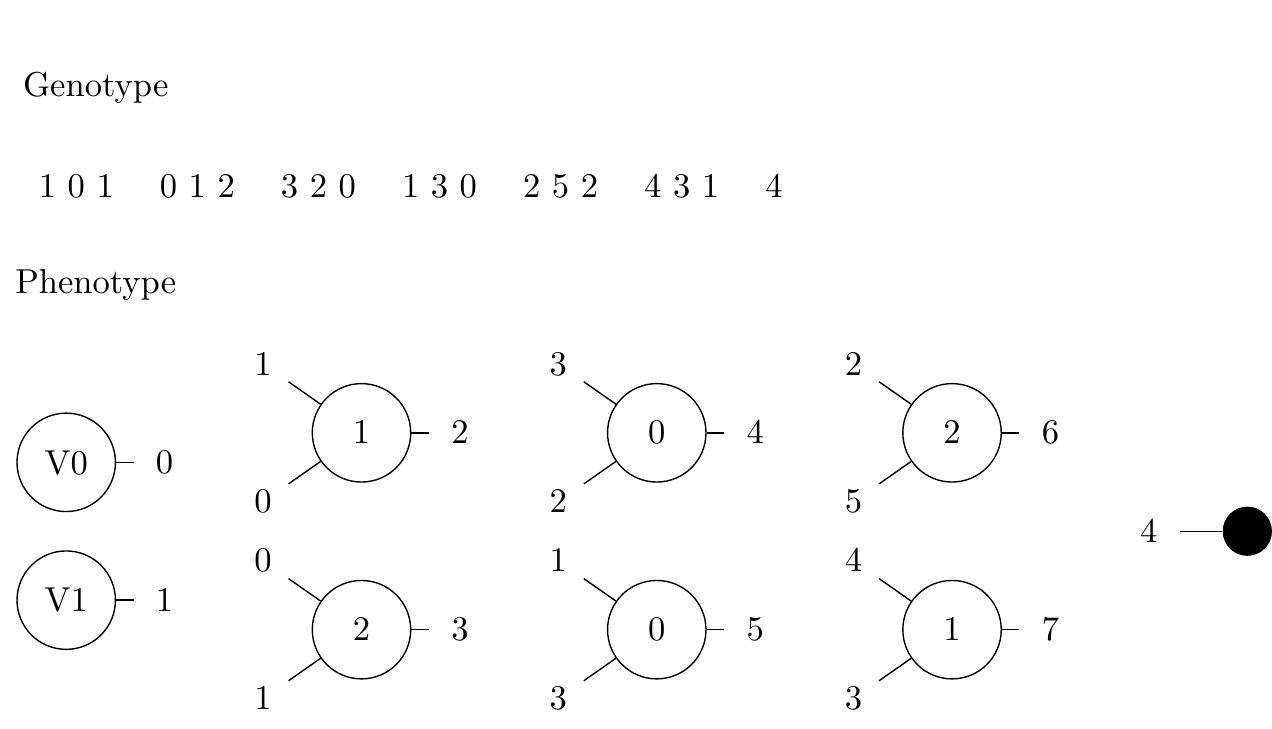}
    \caption{Symbolic representation used by CGP}
    \label{fig:cgprep}
\end{figure}

A related evolutionary algorithm, Cartesian Genetic Programming (CGP), represents the function as a \textit{directed graph}. 
The graph is commonly represented as a two-dimensional grid with given dimensions (number of rows and columns) chosen by the user.
What makes CGP  different from tree-based GP is that in CGP, the genotype is a list of integers representing the graph primitives and their connectivity. 
The genotype is mapped to the directed graph that is evaluated as a Boolean function, as in the case of a GP syntax tree.
CGP genotypes are of fixed length, while the phenotypes have a variable length following the number of unconnected (unexpressed) genes.
Figure~\ref{fig:cgprep} shows how the same expression from Figure~\ref{fig:gprepr} could be represented in CGP. 
The list of integers in the genotype denotes the inputs to each of the nodes (first two numbers in each group) and the function index used by the node (third number in each group).
The function used in each node is determined by mapping the index to one of the available functions, which in this example would be 0 for AND, 1 for OR, and 2 for XOR. 
Finally, the last number in the genotype determines which node represents the output of the expression.
This genotype can be decoded into the illustrated phenotype, in which the nodes are arranged in a graph with two rows and three columns. 
It is interesting to note that not all nodes need to be used when constructing the expression (such is the case with nodes with outputs 5, 6, 7), which enables CGP to evolve expressions of various lengths and complexity. 

It is important to note that symbolic Boolean representations (such as GP and CGP) have been shown to achieve superior results compared to previous encodings in almost all applications.
Symbolic representations have been successfully used to evolve Boolean functions of up to 16 variables~\cite{10.1007/978-3-319-10762-2_41}, which would have been impractical for representations directly encoding a truth table of size $2^{16}$.
Only in the case of specific criteria and smaller Boolean sizes, such as the evolution of hyperbent functions of up to 8 variables~\cite{hyperbent}, may the simple bitstring encoding obtain better results.

\paragraph{Discussion.}
Table~\ref{tab:encodings} provides the division of works that use the truth table encoding (phenotype) considering the genotypes. 
Certain observations can be made: 
\begin{compactenum}
\item The majority of the works use bitstring encoding, which is not surprising as it represents a natural choice. In fact, all the papers between 1997 and 2013 use bitstring encoding. A smaller number of works consider the bitstring encoding restricted to balanced functions, to reduce the size of the search space.
\item Symbolic encoding is the second most used one. Interestingly, while stated in related works to be the most successful one, we cannot notice it has been used more than the bitstring encoding in the last few years.
\item Integer encoding is rarely used, and we know only two papers using it: the first one~\cite{picek18a} considers quaternary Boolean functions, so integer encoding is a natural choice. The second one~\cite{manzoni20} explores two different integer representations for balanced Boolean functions, namely the map-of-ones and the zero-length encodings.
\item Floating-point is, similarly to the integer encoding, rarely used. In fact, there is only one paper using it~\cite{PICEK2017320}. Since that paper is also the only one exploring immunological algorithms for the design of Boolean functions, we can state it represents a not sufficiently investigated approach.
\end{compactenum}

\begin{table}
\small
\centering
\caption{The division of papers using different genotypes and the truth table phenotype.}
\label{tab:encodings}
\begin{tabular}{lp{0.65\textwidth}}
\textbf{Encoding} & \textbf{Papers} \\\toprule
    Bitstring (unconstrained)           &     \cite{7744197, DBLP:conf/ppsn/PicekMBJ14, JAKOBOVIC2021107327, burnett, 10.1007/978-3-319-26841-5_6, 00190, PICEK2016635, hyperbent, PICEK2017320, 10.1145/2739482.2764674, 10.1145/2464576.2464671, 10.1145/2739480.2754739, 10.1145/2739480.2754764, 10.1007/978-3-319-16501-1_16, 10.1007/BFb0028471, 10.1007/10718964_20, 10.1007/3-540-36231-2_20, 10.1007/3-540-48970-3_1, 10.1145/1276958.1277112, KAVUT2010341, Maitra2007BalancedBF, kavutthesis,
DBLP:journals/dcc/KavutMT19, doi:10.1137/16M1107826, 4167738, cryptoeprint:2006/181, millan1997smart, IzbenkoKK09, Lopez-LopezGSOR20, ClarkJS04, FullerDM03, moskovchenko2018heuristic, wang2022searching, BeheraG22}
            \\\midrule
    Bitstring (balanced)             & \cite{10.1007/BFb0054148,10.1007/978-3-319-10762-2_80,manzoni20,ManzoniMT21,ManzoniMT22,10.1109/CEC55065.2022.9870427,MandujanoKL22} \\
    \midrule
     Integer             &      \cite{picek18a,manzoni20}
           \\\midrule
    Floating-point              &        \cite{PICEK2017320}
         \\\midrule
        Symbolic          &        \cite{DBLP:conf/ppsn/PicekMBJ14, 10.1109/CEC55065.2022.9870427, 10.1007/978-3-319-26841-5_6, 00190, PICEK2016635, hyperbent, picek18a, 10.1145/3319619.3321925, 10.1145/2464576.2464671, 10.1007/978-3-319-10762-2_41, 10.1145/2739480.2754764, 10.1007/978-3-319-16501-1_16, 10.1007/978-3-319-10762-2_80}         \\\bottomrule
\end{tabular}
\end{table}

\subsubsection{Criteria}

A natural direction to consider is the number of criteria used in the objective function, which is the function being optimized, while the fitness function is what is used to guide the optimization algorithm. For example, when evolving Boolean functions with high nonlinearity, the common criterion in the objective function is the nonlinearity property. At the same time, to evolve such Boolean functions with high nonlinearity, the fitness function can be of different forms, e.g., using only the nonlinearity property or nonlinearity and the Walsh-Hadamard spectrum.

We divide the works into those that consider only one criterion and those that consider more than one criterion. In the latter case, we will differentiate between the works using two criteria or more than two criteria.

\paragraph{One criterion}

In cases where only one criterion is considered in the optimization process, most of the related works consider nonlinearity. As such, those works usually consider evolving bent Boolean functions, e.g.,~\cite{kavutthesis,FullerDM03,10.1007/978-3-319-10762-2_41}.
Additionally, there is a number of works that aim at balanced Boolean functions with high nonlinearity. In this case, the algorithms usually constrain the search to use only balanced Boolean functions (common examples are local search algorithms that make pairwise bit flips in the truth table) and still maximize nonlinearity as a single criterion in the objective function~\cite{manzoni20,ManzoniMT21,ManzoniMT22}.
From the performance perspective, the employed techniques generally perform well and find bent Boolean functions up to a reasonable number of inputs, e.g., 16. For larger Boolean function sizes, the bottleneck is the evaluation part, as metaheuristics commonly work by assessing a large number of candidate solutions. 
We also mention some previously unattainable results achieved in this category, see, e.g.,~\cite{KAVUT2010341}. Still, we note that the objective function considered nonlinearity only, while success with a specific combination of cryptographic properties cannot be attributed to metaheuristics (since other properties were not optimized but only evaluated afterward).
One example of the unsuccessful use of metaheuristics is the evolution of hyperbent Boolean functions where the criterion is nonlinearity~\cite{hyperbent}.

\paragraph{Two criteria}

With two criteria in the optimization process, there are several variants that need to be discussed.
The simplest option (and the one considered in most of the representative works) uses the nonlinearity~\footnote{The nonlinearity property could be evaluated with only the extreme value in the Walsh-Hadamard transform~\cite{10.1145/2464576.2464671} or as a more elaborate function evaluating the whole Walsh-Hadamard spectrum~\cite{ClarkJS04}.} and balancedness criteria~\footnote{Balancedness appears either as a constraint for a function to have or as an imbalancedness penalty.}. 
Other common options for two criteria include considering 1) nonlinearity and algebraic degree, e.g.,~\cite{DBLP:conf/ppsn/PicekMBJ14}, 2) support and correlation immunity, e.g.,~\cite{10.1007/978-3-319-26841-5_6}, and 3) nonlinearity and autocorrelation, e.g.,~\cite{10.1145/2739482.2764674}.
Finally, we note that there are different ways how the objective function is built for two criteria: some works consider optimizing both criteria at the same time. In contrast, other works employ a two-stage approach where the second criterion is optimized only after the first one is fulfilled.
Moreover, the related works could also be divided based on whether they consider maximizing (minimizing) both criteria or minimizing one criterion and maximizing the other.
As is the case for one criterion, optimizing for two criteria seems to be a relatively easy problem where the bottleneck becomes the Boolean function size. One example of a successful result would be finding balanced Boolean functions with 8 inputs and nonlinearity equal to 116~\cite{PICEK2016635}, which is also the best-known result in the literature.

\paragraph{More than two criteria}

There are multiple options for the objective function when considering more than two criteria. In fact, our analysis shows most works in this category consider three or four criteria, but there are works going up to six criteria.
Some common combinations include 1) nonlinearity, algebraic degree, and balancedness~\cite{00190}, and 2) nonlinearity, autocorrelation, and balancedness~\cite{10.1007/3-540-48970-3_1}.
On the other hand, unusual combinations include 1) nonlinearity, algebraic immunity, and fast algebraic resistance~\cite{cryptoeprint:2013:011}, 2) nonlinearity, correlation immunity, and strict avalanche criterion (SAC)~\cite{10.1007/BFb0054148,10.1145/2739482.2764674}, and 3) balancedness, nonlinearity, and transparency order~\cite{10.1007/978-3-319-10762-2_80}.
Besides the approaches discussed in the previous paragraph (optimizing multiple criteria at the same time and multi-stage approach), here, we also notice several works using the multiobjective paradigm~\cite{doi:10.1137/16M1107826,10.1145/1276958.1277112}. 
From the performance perspective, the results achieved in this category are mostly good, where, as previously, the bottleneck becomes the Boolean function size. Additionally, in this category, we can also recognize the bottleneck due to the slow computation of specific cryptographic properties (e.g., algebraic immunity)~\cite{doi:10.1137/16M1107826}.

\subsubsection{Boolean Function Size}

From the perspective of optimizing Boolean functions of different sizes, we can observe several trends. First, the smallest considered size is Boolean functions with four inputs, see, e.g.,~\cite{10.1145/2739480.2754739,10.1145/1276958.1277112}.\footnote{Actually, for the quaternary Boolean functions, the authors~\cite{picek18a} start with dimension two, but once mapped to binary Boolean functions, it is naturally larger.}
Next, most of the works consider smaller sizes, e.g., up to eight inputs, with eight inputs being the most investigated dimension in general~\cite{DBLP:conf/ppsn/PicekMBJ14,burnett,PICEK2017320,millan1997smart}. The works that consider larger sizes commonly go up to 16 inputs~\cite{00190,PICEK2017320,10.1007/978-3-319-10762-2_41,10.1007/BFb0028471,FullerDM03,cryptoeprint:2013:011}. We are aware of only one paper that considers larger than 16 inputs Boolean functions where the authors evaluate their approach up to 26 inputs~\cite{DBLP:journals/dcc/KavutMT19}.

\subsection{Walsh-Hadamard-based Representation}
\label{subsec:whs}

We have seen in Section~\ref{subsec:prop} that the Walsh-Hadamard spectrum can be used to characterize several cryptographic criteria of a Boolean function, including its balancedness, nonlinearity, and correlation immunity order. Thus, an interesting idea is to encode a candidate solution as a Walsh-Hadamard spectrum that already satisfies certain properties. At this point, one might think it is possible to manipulate this spectrum using a metaheuristic and optimize for other properties not captured by the Walsh-Hadamard transform (for example, the algebraic degree). However, the situation is more complicated: as we mentioned in Section~\ref{subsec:bf-rep}, the WHT is an injective mapping from $\F_2^{2^n}$ to $\mathbb{Z}^{2^n}$, but clearly not a surjective one. Hence, if one starts from a random spectrum and then applies the inverse transform, the result likely is a \textit{pseudo-Boolean function} $f: \F_2^n \to \mathbb{Z}$, rather than a Boolean function. This suggests the following strategy to use the Walsh-Hadamard spectrum as a representation method for metaheuristics:
\begin{compactenum}
    \item Encode the genotype of the candidate solution as a WHS satisfying the desired set of properties, e.g., low maximum absolute value for high nonlinearity and coefficients up to Hamming weight $t$ set to 0 for correlation immunity of order $t$.
    \item Evaluate the fitness of this candidate solution by applying the inverse Walsh-Hadamard transform and then measuring how far the obtained pseudo-Boolean function is from being a true Boolean function.
    \item Apply the variation operators of the metaheuristic to modify the WHS of the candidate solution, steering the corresponding pseudo-Boolean function closer to a true Boolean function.
\end{compactenum}

The approach above, called \textit{spectral inversion}, was pioneered by Clark et al. in~\cite{1299941}. There, the authors focused on the search for plateaued Boolean functions. In particular, a Boolean function $f: \F_2^{n} \to \F_2$ is called plateaued if its Walsh-Hadamard coefficients are at most three-valued, and namely, they range in the set $\{-2^r, 0, +2^r\}$, with $r \le \frac{n}{2}$. The case $r = \frac{n}{2}$ actually corresponds to bent functions, which can be seen as a subset of plateaued functions characterized by spectra that only take two values in $\{-2^{\frac{n}{2}}, +2^{\frac{n}{2}}\}$. When $r$ is strictly larger than $\frac{n}{2}$, the spectrum of a plateaued function must have some coefficients set to 0 due to Parseval identity. Therefore, proper plateaued functions are interesting because they can be balanced and correlation immune. Is possible to prove (see, e.g.,~\cite{carlet_2021}) that a plateaued function $f: \F_2^n \to \F_2$ with $r > \frac{n}{2}$ has resiliency order $r-2$ (thus it is balanced and correlation immune of order $r-2$), nonlinearity $2^{n-1}-2^{r-1}$, and algebraic degree $n-r-3$. Consequently, it is optimal both concerning Siegenthaler's and Sarkar and Maitra's bounds.

The idea behind the representation of Clark et al.'s spectral inversion method is the following. Once the target $n$ and $r$ are chosen, a candidate solution is encoded as a three-valued Walsh-Hadamard spectrum. The coefficients corresponding to positions with Hamming weight up to $r-2$ are always set to 0 to ensure resiliency order $r-2$. Then, theoretical results, including Parseval's identity and Sarkar-Maitra's divisibility bound, are used to determine the number of remaining coefficients that need to be set respectively to $-2^r$, $0$, and $+2^r$. These remaining coefficients can be freely permuted in the positions of the spectrum with Hamming weight higher than $r-2$.

To evaluate the fitness of a three-valued spectrum, Clark et al. experimented with two fitness functions, both measuring the distance of the pseudo-Boolean function obtained by applying the inverse WHT to the spectrum from being a true Boolean function. In particular, a global optimum for this problem (i.e., reaching distance zero) corresponds to a plateaued Boolean function with the desired profile of properties encoded by its spectrum. As a metaheuristic to drive the search, the authors used a simulated annealing algorithm. The simulated annealing algorithm's basic move consisted of swapping two distinct values in the spectrum (excluding the coefficients set to zero for resiliency). The authors applied this method to evolve plateaued functions of size 7, 8, and 9, but with a very low success rate only for 7 variables, while no functions were produced for 8 and 9 inputs.

The principle of spectral inversion has been investigated in several more works, using other metaheuristics or focusing on subclasses of plateaued functions. For example, Stanica et al. used again simulated annealing to evolve Rotation-Symmetric Boolean Functions (RSBF), where the input vectors that are equivalent under cyclic rotations have the same output values~\cite{StanicaMC04}. Reducing the search space in this way, the authors were able to construct $9$-variable plateaued functions with nonlinearity 240, resiliency order 2, and degree 6.

Saber et al. used Particle Swarm Optimization to obtain a $9$-variable plateaued function with nonlinearity 240, resiliency order 3, and algebraic degree 5~\cite{saber06}. The PSO version used in that work~\footnote{The details of the PSO algorithm are available in the master thesis~\cite{uddin06} of the second author of~\cite{saber06}.} is quite different from the traditional one used to solve continuous optimization problems. In particular, this is a discrete version of PSO suitable for permutation or combinatorial search spaces. Updating a particle's position amounts to swapping two distinct values in the Walsh-Hadamard spectrum, much like in the simulated annealing approach of~\cite{1299941}. Velocity vectors are instead replaced by probability vectors: the higher the velocity of a particle along a specific coordinate, the more likely the corresponding spectral value is swapped with another one.~\footnote{The PSO algorithm of~\cite{10.1145/2739482.2764674} surveyed in Section~\ref{subsec:tt} uses the same principle to optimize balanced Boolean functions with a truth table-based encoding, with the swap-based position update used to maintain the balancedness.} 

Kavut et al. devised a steepest-descent algorithm to search for $9$-variable plateaued functions with the same properties of those considered in~\cite{saber06} through spectral inversion~\cite{resilient_concatenation_spectra}. This metaheuristic basically corresponds to a greedy local search where the move giving the best improvement is always chosen at each iteration until a local optimum is reached. The success rate of this technique turned out to be quite low (6 functions obtained out of 150 runs), but this was sufficient for the authors' purposes since they used these functions as a basis to construct larger ones via concatenation.

More recently, Mariot and Leporati proposed a genetic algorithm to evolve plateaued functions of 6 and 7 variables using spectral inversion~\cite{10.1007/978-3-319-26841-5_3}. The genotype representation was the same, i.e., a three-valued Walsh-Hadamard spectrum with coefficients set to zero up to Hamming weight $r-2$. To create valid offspring chromosomes, the authors designed ad-hoc crossover and mutation operators that would preserve the properties of the spectra. In particular, the crossover was based on the idea of counters to keep track of the multiplicities of $-2^r$, $0$, and $+2^r$, similar to what is done in the works that evolve balanced Boolean functions with GA~\cite{manzoni20}. The authors of~\cite{10.1007/978-3-319-26841-5_3} obtained good success rates for plateaued functions of $6$ variables, outperforming the simulated annealing algorithm of~\cite{1299941}, but they were not able to produce any plateaued function of $7$ variables.

\section{Metaheuristic-assisted Construction of Boolean Functions}
\label{sec:assisted_construction}

\subsection{Evolving Constructions}

It is common to divide algebraic constructions into primary and secondary ones~\cite{carlet_2021}. In primary constructions, new functions are obtained without using known ones. In secondary constructions, existing functions are used to construct new ones.
Today, we know several constructions that can be used to obtain Boolean functions with specific cryptographic properties.
For example,~\cite{carlet_2021} lists ten primary and ten secondary constructions to obtain bent Boolean functions. While this can be considered as numerous, there is no reason why there could not be many more possible ones. Additionally, bent Boolean functions can also be considered a well-explored topic~\cite{Mesnager2016} compared to some other Boolean functions. Indeed, we could easily envision a specific set of properties that a Boolean function needs to fulfill and for which there is no available algebraic construction.

In such settings, it makes sense to ask whether metaheuristics could find algebraic constructions.
Since most algebraic constructions are given in the symbolic form, they can be optimized using a suitable encoding.
Here, the GP methods offer a natural mapping using either a tree-based or graph-based representation.
Several papers have addressed the topic of evolving, rather than inventing, secondary Boolean constructions; these papers differ regarding the method and the objective criteria.

In~\cite{10.1145/2908812.2908915}, the authors aim at obtaining bent Boolean functions for a larger number of variables.
Rather than directly evolving for bent functions as in, e.g.,~\cite{10.1007/978-3-319-10762-2_41}, the authors used GP to evolve a secondary construction to transform several input bent functions of $n-2$ variables into an output bent function of $n$ variables (inspired by the Rothaus construction).
Using as seeds bent functions of only 4 variables, the authors have obtained secondary constructions that generate bent functions in higher dimensions.
Apart from that, the evolved constructions seem to be quite general because the same constructions have succeeded in generating bent functions from 6 up to 24 variables from different seed functions (in two fewer variables). What remains unclear from the obtained results is whether any of the evolved constructions would be a new one. Indeed, since metaheuristics find numerous solutions, one would need to check all of them to see if there are new constructions. Moreover, as it is not possible to know the size of the construction to be generated, they commonly have a significant amount of bloat that must be analyzed and removed.

Mariot et al. used the approach from the previous paragraph and tried to evolve constructions of hyperbent Boolean functions~\cite{hyperbent}. Unfortunately, the approach did not work, and the authors did not manage to evolve any constructions resulting in hyperbent Boolean functions.

A similar approach is used in~\cite{10.1145/3512290.3528871}, but this time to find balanced functions with high nonlinearity. The approach is again based on finding secondary constructions, which are, in turn, evolved with GP.
This approach may be interesting since only a few known such constructions exist. 
Their results show that GP can find constructions that tend to generalize well, i.e., result in balanced highly-nonlinear functions for various tested sizes and different input function groups.
While the obtained levels of nonlinearity rarely reach the best-known values, the simplest solution found by GP turns out to be a particular case of the well-known indirect sum construction.

Carlet et al. discuss how evolutionary algorithms can be used to help in finding a good Boolean function construction~\cite{DBLP:conf/gecco/CarletJP21}. The authors start from a recent work by C. Carlet, where he proposed a generalization of the Hidden Weight Boolean Function allowing a construction of $n$-variable balanced functions $f$ from $(n-1)$-variable Boolean functions $g$ fulfilling some criteria. There are multiple choices for the function $g$, and the authors used evolutionary algorithms to find functions that still satisfy the necessary criteria while improving nonlinearity. The approach resulted in Boolean functions with significantly improved nonlinearity.

Mariot et al. investigated how to design a secondary semi-bent and bent construction of Boolean functions based on cellular automata~\cite{DBLP:journals/nc/MariotSLM22}. More precisely, the authors started with Boolean functions with good cryptographic properties and used them as local rules of cellular automata to obtain larger Boolean functions with similar cryptographic properties. The choice of the local rule was performed using an evolutionary algorithm, which evolved an affine transformation that preserved the main cryptographic properties of the starting function (e.g., nonlinearity).

\subsection{Construction of Related Objects}
\label{subsec:related}

In this last section, we survey a few works focused on using metaheuristics to construct combinatorial objects related to Boolean functions. Although most of these works are not directly motivated by the search for Boolean functions with good cryptographic properties, we show how some of them can be interpreted as metaheuristic constructions of particular classes of Boolean functions.

The field of combinatorial designs concerns the study of families of subsets of a finite set, such that they satisfy certain balancedness properties~\cite{stinson04}. Besides being a source of interesting open problems in discrete mathematics, the interest in combinatorial designs also spawns from the multiple applications they have in diverse domains, including statistics, the design of experiments, error-correcting codes, and cryptography.

Similarly to the case of Boolean functions, most of the constructions of combinatorial designs proposed in the literature leverage the use of algebraic methods~\cite{colbourn10}. An interesting research thread that emerged in the last few years also considers the metaheuristic construction of specific combinatorial designs. These include the use of various optimization algorithms such as simulated annealing and evolutionary algorithms to construct orthogonal arrays~\cite{safadi1992,wang1992,MariotPJL18,manzoni20}, Steiner systems~\cite{ashlock1996}, orthogonal Latin squares~\cite{MariotPJL17}, disjunct matrices~\cite{KnezevicPMJL18}, and permutation codes~\cite{Mariot2022}.

Some classes of Boolean functions with good cryptographic properties can be characterized in terms of combinatorial designs. For example, bent functions are equivalent to Hadamard matrices and difference sets of a specific form~\cite{Rothaus,dillon74}. More relevant to our discussion is the fact that correlation immune functions can be characterized in terms of orthogonal arrays. An orthogonal array of $N$ runs, $k$ levels, $s$ entries, and strength $t$, denoted as an $OA(N,k,s,t)$, is a $N\times k$ array with entries from a set of $s$ symbols, such that in every $N \times t$ subarray, each $t$-uple of symbols occurs exactly $\lambda = N/s^t$ times. Binary orthogonal arrays are OAs with $s=2$. The connection between Boolean functions and binary orthogonal arrays stems from a result by Camion et al.~\cite{CamionCCS91}, where they proved that a Boolean function $f: \F_2^n \to \F_2$ is $t$-th order correlation immune if and only if its support is an $OA(N,k,2,t)$, where $N = \#supp(f)$ and $k=n$. Therefore, works using optimization algorithms for constructing binary orthogonal arrays can be equally interpreted as a metaheuristic construction of $t$-th order correlation immune functions.

To the best of our knowledge, only two works explicitly address the construction of binary orthogonal arrays, namely~\cite{MariotPJL18} and~\cite{manzoni20}. In the former, the authors use genetic algorithms and genetic programming to evolve $N\times k$ binary matrices. Given the target parameter $t$, the fitness function measures the deviation of each $N\times t$ submatrix from satisfying the balancedness constraint required for an OA of strength $t$. For the GA variation operators, the counter-based balanced crossover of~\cite{10.1007/BFb0054148} and swap mutation are applied column-wise on the binary matrices in the population. This is based on the observation that any $OA(N,k,s,t)$ is also an $OA(N,k,s,i)$ for all strength $i$ up to $t-1$. Thus, any binary OA must also have balanced columns. This idea has been further explored in~\cite{manzoni20} with the map-of-ones and zero-length balanced crossover operators. The GP algorithm used in~\cite{MariotPJL18} is based instead on a representation similar to the symbolic encoding surveyed in Section~\ref{subsec:tt}. A $N\times k$ matrix is encoded by a set of $k$ syntactic trees that are used to synthesize the columns of the matrix. In particular, a column is the truth table vector obtained by evaluating the corresponding tree over all possible inputs. Consequently, the number of runs $N$ of the array is forced to be a power of $2$ since it is basically the truth table of a Boolean function.

The results of~\cite{MariotPJL18} showed that GP generally outperforms GA on this particular problem, although GA effectively searches a smaller search space due to the representation with balanced columns. In particular, GP can construct orthogonal arrays up to $OA(32, 16, 2, 3)$ and $OA(32,31,2,2)$, while GA arrives at most at $N=16$ runs. The number of runs of the OA is particularly interesting when considering correlation immune functions as a countermeasure for side-channel attacks. In this scenario, it is desirable that the support of such functions is as low as possible for efficiency reasons. In the OA interpretation of correlation immune functions, the support size corresponds to the number of runs $N$ of the arrays.

The shortcoming of~\cite{MariotPJL18} is that the number of runs must be set before the evolutionary search begins. However, one could envision a two-stage optimization process where in the first step, one constructs an OA with specified parameters set as done in~\cite{MariotPJL18}. Then, one can consider the reduction of the number of runs as a combinatorial optimization problem itself. The idea is to select a subset of $2^t$ rows so that the reduced matrix is still an OA with a $\lambda$ parameter decreased by 1. The choice of this subset of rows can be made again using a metaheuristic algorithm. This method has been investigated in~\cite{DBLP:journals/corr/abs-2111-13047}, where the author devised a GA to perform this reduction step. However, the results reported there are only for very small OAs, and this approach should be investigated more in the future.

\section{Potential Research Directions}
\label{sec:directions}

While the existing works consider different metaheuristic techniques and objectives when constructing Boolean functions with good cryptographic properties, we can still identify several future research directions.
\begin{compactenum}
\item Using truth table solution encoding, especially in combination with bitstring representation, represents a standard and well-explored option. We do not see significant opportunities there besides considering previously not investigated combinations of cryptographic properties. 
\item Floating-point representation deserves further investigation to assess what advantages it can bring (if any).
\item Truth table encoding is a natural option but will always suffer from the computational bottleneck and cannot be used to construct large Boolean functions.
\item We found no works using ANF solution encoding. Since many bounds and properties can be expressed through it, it would be interesting to use ANF to encode solutions.
\item WHS encoding is interesting, but it suffers from solutions that do not map to Boolean functions. This problem is especially pronounced in the context of operators (like mutation and crossover) as they easily disrupt a correct solution into a wrong one. It remains an open question of how to construct metaheuristic operators that can work on WHS encoding and maintain the correctness of solutions.
\item In the same way that researchers constructed metaheuristic operators that preserve balancedness for the truth table encoding, it would be relevant to explore how to construct analogous operators for symbolic encoding.
\item We consider metaheuristic-assisted approaches to be the option of choice for designing Boolean functions of arbitrary sizes. Since only a few works are exploring it, more work is needed. For instance, metaheuristic approaches to obtaining algebraic constructions are relevant, but we are missing identifying what constructions are needed. One option would be to find new constructions to serve as an alternative to already-known ones (e.g., bent Boolean functions constructions). Another option would be to identify relevant properties of Boolean functions that cannot be achieved with the known constructions and try to design such constructions.
\item Finally, constructing related objects that can be later transformed into Boolean functions with desired properties is an interesting but unexplored domain. The first step is identifying potentially interesting related objects and assessing if constructing such objects with metaheuristics is easier than directly constructing Boolean functions. Ideas of possible combinatorial designs to evolve with metaheuristics are partial spreads and Hadamard matrices, which are related to bent functions.
\end{compactenum}

\section{Conclusions}
\label{sec:conclusions}

In this survey, we provide an analysis of works that use metaheuristics to construct Boolean functions with good cryptographic properties.
We provide a new taxonomy based on the solution encoding rather than the search technique used since many of the related works actually use combinations of construction techniques, making them difficult to be classified. On the other hand, when considering the solution representation, we can see that works naturally map into different categories, allowing easier analysis of the results. Finally, we identified and discussed potentially interesting directions for future research.

\bibliographystyle{abbrv}
\bibliography{bibliography}

\end{document}